\documentclass[a4paper,oneside]{scrartcl}

\usepackage{enumerate}
\usepackage[english]{babel}
\usepackage{amsthm}
\usepackage{amssymb}
\usepackage{amsmath}
\usepackage{amsfonts}
\usepackage[footnote,marginclue]{fixme}
\usepackage{xspace}
\usepackage{xcolor}
\usepackage[capitalize]{cleveref}

\theoremstyle{plain}
\newtheorem{theorem}{Theorem}[section]
\newtheorem{lemma}[theorem]{Lemma}

\newtheorem{proposition}[theorem]{Proposition}

\theoremstyle{definition}

\newtheorem{definition}[theorem]{Definition}

\newcommand{\rel}{\mathrm{rel}}
\newcommand{\mA}{{\mathfrak A}}

\newcommand{\dom}{{\rm dom}}

\newcommand{\FO}{{\rm FO}}

\newcommand{\Fr}{{\rm Fr}}
\newcommand{\Var}{{\rm Var}}
\newcommand{\SO}{{\rm SO}}
\newcommand{\ESO}{{\rm ESO}}
\def\dep{=\!\!}
\newcommand{\df}{\mathcal{D}}
\newcommand{\most}{{\sf Most}}
\newcommand\func{_{\mathrm{f}}}
\newcommand{\mostf}{{\sf Most\func}}
\newcommand{\rc}{{\sf R}}
\newcommand{\M}{{\sf M}}

\newcommand{\dfn}{\mathrel{\mathop:}=}

\newcommand{\np}{{\rm NP}}

\newcommand{\ch}{{\rm CH}}
\newcommand{\p}{{\rm P}}
\newcommand{\pp}{{\rm PP}}
\DeclareMathOperator{\CkP}{C_{\mathit k}P}
\DeclareMathOperator{\ClP}{{C_{\mathit l}P}}
\DeclareMathOperator{\CP}{C_0P}

\DeclareMathOperator{\Ck1P}{C_{\mathit k+1}P}

\newcommand{\tu}[1]{\overline{#1}}

\definecolor{darkgreen}{rgb}{0,.7,0}
\definecolor{darkyellow}{rgb}{.7,0.7,0}
\newcommand{\note}[2][]{\textnormal{\color{gray}\scriptsize++\ifthenelse{\equal{#1}{}}{}{#1 says: }#2++}}
\newcommand{\todo}[2][]{\textnormal{\color{red}\scriptsize+++\ifthenelse{\equal{#1}{}}{}{#1 says: }#2+++}}

\begin{document}
\author{Arnaud Durand\thanks{Universit\'e Paris Diderot, IMJ, CNRS UMR 7586, Case 7012, 75205 Paris cedex 13, France, durand@logique.jussieu.fr}, 
Johannes Ebbing\thanks{Leibniz Universit\"at Hannover, Theoretical Computer Science, Appelstr.~4, 30167~Hannover, Germany, \{ebbing,vollmer\}@thi.uni-hannover.de}, 
Juha Kontinen\thanks{University of Helsinki, Department of Mathematics and Statistics,
P.O. Box 68, 00014, Finland, juha.kontinen@helsinki.fi.}, 
Heribert Vollmer\footnotemark[3]}

\title{Dependence logic with a majority quantifier\footnote{The third author was supported by grants 127661 and 138163 of the Academy of Finland. The second and fourth author were supported by a grant from DAAD within the PPP programme. The fourth author was also supported by DFG grant VO 630/6-2.}}

\maketitle

\begin{abstract}
We study the extension of dependence logic $\df$ by a majority quantifier $\M$ over finite structures. We show that the resulting logic is equi-expressive with the extension of second-order logic by second-order majority quantifiers of all arities. Our results imply that, from the point of view of descriptive complexity theory, $\df(\M)$  captures the complexity class  counting hierarchy.  

\end{abstract}

\section{Introduction}
We study the extension of dependence logic $\df$  by a majority quantifier $\M$ over finite structures. Dependence logic \cite{vaananen2007}  extends first-order logic by  dependence atomic formulas 
\begin{equation*}\dep(t_1,\ldots,t_n)
\end{equation*} the intuitive meaning of which is that the value of the term $t_n$ is completely determined by the values of $t_1,\ldots, t_{n-1}$. While in first-order logic the order of quantifiers solely determines the dependence relations between variables, in dependence logic more general dependencies between variables can be expressed. Historically dependence logic was preceded by  partially ordered quantifiers (Henkin quantifiers) of Henkin  \cite{henkin1961} and Independence-Friendly (IF) logic of Hintikka and Sandu \cite{MR1034575}. It is known that both IF logic and dependence logic are equivalent to existential second-order logic $\ESO$ in expressive power.
From the point of view of descriptive complexity theory, this means that dependence logic captures the class $\np$.

The framework of dependence logic has turned out be flexible to allow  
interesting generalizations. For example, the extensions of dependence logic in terms of so-called intuitionistic implication and linear implication was  introduced in \cite{AbVa}. In \cite{YangUNPUBLISHED} it was shown that extending $\df$ by the intuitionistic implication makes the logic equivalent to full second-order logic $\SO$. 

Recently, new variants of the dependence atomic formulas have been introduced in \cite{JoukoUNPUBLISHED} and \cite{PietroUNPUBLISHED}. Also a modal version of dependence logic was introduced in \cite{MDLJouko} and has  been studied in \cite{LV10} and \cite{Sevenster09}. In this paper we are concerned with introducing a new quantifier to dependence logic: the majority quantifier.
Adding majority and, more generally, counting capabilities to logical formalisms or computational devices has deserved a lot of attention in theoretical computer science. Understanding the power of counting is an important problem both in logic and  in computational complexity:
\begin{itemize}
\item The circuit class TC$^{0}$, the class of problems solvable by polynomial-size constant-depth circuits with majority gates, is at the current frontier for lower bound techniques (see, e.g., \cite{vol99}). We have strict separations of classes within TC$^{0}$, but above TC$^{0}$ we have essentially no lower bounds. By a diagonalization it follows that TC$^{0}$ is different from the second level of the exponential-time hierarchy, but a separation from a lower class seems to be far away. In particular, the question if TC$^{0}$ equals NC$^{1}$ (logarithmic-depth circuits with bounded fan-in gates) is considered the P-NP problem of circuit complexity.
(Concerning uniform circuits, we know that uniform TC$^{0}$ is strictly included in the class PP of probabilistic polynomial time \cite{MR1745163}.)
\item The counting-hierarchy (the oracle hierarchy built upon PP) can be characterized using majority quantifiers in just the same way as by Wrathall's theorem existential and universal quantifiers characterize the polynomial hierarchy \cite{Toran:Comcdc}.
\item By Toda's theorem, one majority quantifier is as powerful as the whole polynomial hierarchy \cite{Toda:PPiah}.
\end{itemize}

Here we suggest a definition of a majority quantifier for dependence logic. The proposed semantics mimics that of the existential and universal quantifiers in $\df$. The present paper is devoted to a first study of the resulting logic, denoted by $\df(\M)$. We examine some of its basic properties, prove strong normal forms (some of our technically most involved proofs are found here), and show in our main result, that dependence logic with the majority quantifier leads to a new descriptive complexity characterization of the counting hierarchy: $\df(\M)$ captures $\ch$.

Engstr\"om \cite{FredrikUNPUBLISHED} has also studied generalized quantifiers in dependence logic. He considered different conservative extensions of $\df$---informally this means that he extends $\df$ by generalized quantifiers in a first-order manner. From a descriptive complexity point of view, his logics do not lead out of $\np$, i.e., $\ESO$, assuming  the quantifier in question is $\ESO$-definable (e.g., the majority quantifier).
Our approach and results differ from that of Engstr\"om since we are in a sense extending dependence logic by a dependence majority quantifier, whose semantics is defined in close analogy with the semantics of $\exists$ and $\forall $ in dependence logic. 
The results of our paper show that our extension behaves like an extension of $\SO$ by second-order generalized  quantifiers.

This article is organized as follows. In Section 2 we defined dependence logic and discuss some basic results on it. Then we introduce a majority quantifier for the dependence logic setting and discuss the basic properties of $\df(\M)$. In subsection 2.3 we discuss the complexity  class counting hierarchy and the second-order majority quantifiers $\most^k$ that have been used to characterize it in \cite{KontinenJ:Logcch}. In Section 3, we introduce second-order majority quantifiers $\most^k\func$ ranging over functions and in Section 4 we show that, for sentences  the logics $\SO(\mostf)$ (the extension of second-order logic $\SO$ by  $\most^k\func$ for $k\ge1$) and $\df(\M)$ are equivalent.

\section{Preliminaries}

In this section we first define dependence logic and discuss  its basic properties. Then we define the counting hierarchy and the logic corresponding to it.

\subsection{Dependence Logic}

Dependence logic ($\df$) extends the syntax of first-order logic by new dependence atomic formulas. In this article we consider only formulas of $\df$ that are in negation normal form.
\begin{definition}[\cite{vaananen2007}] Let $\tau$ be a  vocabulary.
The $\tau$-formulas of dependence logic  ($\df[\tau]$) is defined by extending $\FO[\tau]$, defined in terms
of $\vee$, $\wedge$, $\neg$, $\exists$ and $\forall$, by atomic dependence formulas 
\begin{equation}\label{dep}\dep(t_1,\ldots,t_n),
\end{equation} where $t_1,\ldots,t_n$ are terms. 
\end{definition}
The  meaning of the formula \eqref{dep} is
that the value of the term $t_n$ is functionally determined by the values of
the terms $t_1,\ldots, t_{n-1}$.  The formula $\dep()$ is interpreted as $\top$. The semantics of $\df$ will be formally presented shortly.
\begin{definition} Let $\phi\in \df$. The set  $\Fr(\phi)$  of free variables of a formula $\phi$
is defined  as for first-order logic, except that we have the new case
\[ \Fr(\dep(t_1,\ldots,t_n))=\Var(t_1)\cup\cdots \cup \Var(t_n), \]
where $\Var(t_i)$ is the set of variables occurring in term $t_i$. If $\Fr(\phi)=\emptyset$,
we call $\phi$ a sentence.
\end{definition}

The semantics of $\df$ is formulated using the concept of a \emph{Team}. Let $\mA$ be a model with domain $A$.
 {\em Assignments} of $\mA$
are finite mappings from variables into $A$. The value of a term $t$
in an assignment $s$ is denoted by $t^{\mA}\langle s\rangle$.
If $s$ is an assignment, $x$ a variable,  and $a\in A$, then $s(a/x)$ denotes the
assignment (with domain $\dom(s)\cup \{x\}$)  that agrees with $s$
everywhere except that it maps $x$ to $a$.

\begin{definition}
Let $A$ be a set and $\{x_1,\ldots,x_k\}$ a finite (possibly empty) set  of
variables. 
\begin{enumerate}
\item  A {\em team} $X$ of $A$ with domain
$\dom(X)=\{x_1,\ldots,x_k\}$ (we call $A$ the \emph{co-domain} of $X$) is any set of assignments  $s\colon \{x_1,\ldots,x_k\}\rightarrow A$.
\item The relation $\rel(X)\subseteq A^k$ corresponding to $X$  is defined as
\[\rel(X)=\{(s(x_1),\ldots,s(x_k)) : s\in X \}.  \]
\item For a function $F\colon X\rightarrow A$, we define
\begin{eqnarray*}
X(F/x)&=&\{s(F(s)/x) : s\in X \}\\ 
 X(A/x)&=&\{s (a/x): s\in X\ \textrm{and}\ a\in A \}.
\end{eqnarray*}

\end{enumerate}
\end{definition}
We will next define the semantics of dependence logic. Below,  atomic formulas and their negations are called literals.
\begin{definition}[\cite{vaananen2007}]\label{sat} Let
$\mA$ be a model and $X$ a team of $A$. The satisfaction relation
$\mA\models _X \phi$ is defined as follows:
\begin{enumerate}
\item If $\phi$ is a first-order literal, then $\mA\models _X\phi$ iff for all $s\in X$ we have  $\mA\models_s\phi$. 

\item $\mA\models _X \dep(t_{1},\ldots,t_{n})$ iff for all $s,s'\in
X$ such that\\ $t_1^{\mA}\langle s\rangle  =t_1^{\mA}\langle
s'\rangle  ,\ldots, t_{n-1}^{\mA}\langle s\rangle
=t_{n-1}^{\mA}\langle s'\rangle  $, we have $t_n^{\mA}\langle
s\rangle  =t_n^{\mA}\langle s'\rangle  $.

\item  $\mA \models _X \neg \dep(t_{1},\ldots,t_{n})$ iff
$X=\emptyset$.

\item $\mA\models _X \psi \wedge \phi$ iff $\mA\models _X \psi$ and $\mA\models _X \phi$.
\item $\mA\models _X \psi \vee \phi$ iff $X=Y\cup Z$ such that
 $\mA\models _Y \psi$  and $\mA\models _Z \phi$ .

\item   $\mA \models _X \exists x\psi$ iff $\mA \models _{X(F/x)} \psi$ for some $F\colon X\to A$.

\item $\mA \models _X \forall x\psi$ iff $\mA \models _{X(A/x)} \psi$.

\end{enumerate}
Above, we assume that the domain of $X$ contains the variables free in $\phi$. Finally, a sentence $\phi$ is true in a model $\mA$  (in symbols: $\mA\models \phi$)  if $\mA\models _{\{\emptyset\}} \phi$. Above, $A \models_s \phi$ denotes satisfaction in first-order logic.
\end{definition}

Let us then  recall some basic properties of dependence logic that will be needed later. The following lemma shows that
the truth of a $\df$-formula depends only on the interpretations of variables occurring free in the formula. Below,  for $V\subseteq \dom(X)$, $X\upharpoonright V$ is defined by
 \[X\upharpoonright V \dfn\{s\upharpoonright V \mid s\in X\}.\]
\begin{lemma}[\cite{vaananen2007}]\label{freevar}
Suppose $V\supseteq \Fr(\phi)$. Then $\mA \models _X\phi$ if and only if $\mA \models _{X\upharpoonright V} \phi$.
\end{lemma}

All formulas of dependence logic also satisfy the following strong monotonicity property called  \emph{Downward Closure}.
\begin{proposition}[\cite{vaananen2007}]\label{Downward closure}Let $\phi$ be a formula of dependence logic, $\mA$  a model, and $Y\subseteq X$ teams. Then $\mA\models_X \phi$ implies $\mA\models_Y\phi$.
\end{proposition}
On the other hand, the expressive
power of sentences of $\df$ coincides with that of existential
second-order sentences:
\begin{theorem}[\cite{vaananen2007}]\label{ew}$\df= \ESO$.
\end{theorem}
Finally, we note that  dependence logic is a conservative extension of first-order logic.
\begin{definition} A formula $\phi$ of $\df$ is called a first-order formula if it does not contain dependence atomic formulas as subformulas.  \end{definition}

First-order formulas of dependence logic satisfy the so-called \textit{flatness} property:
\begin{theorem}[\cite{vaananen2007}]\label{FO} Let $\phi$ be a first-order formula of dependence logic. Then for all $\mA$ and $X$:
\[ \mA\models _X\phi \textrm{ if and only if for all $s\in X$ we have } \mA\models_s\phi.\]
\end{theorem}

\subsection{Dependence logic with a majority quantifier}
The main topic of the present paper is the study of a logic obtained from $\df$ by the introduction of a majority quantifier $\M$. We denote this extended logic by $\df(\M)$. 
It is formally defined by extending the  syntax and semantics of dependence logic by the following clause:
\begin{center}
 $\mA\models _X \M x\phi(x)$ iff for at least $|A|^{|X|}/2$ many functions $F\colon X\rightarrow A$ we have $\mA\models_{X(F/x)}\phi(x)$.
\end{center}
Analogously to $\df$ the logic $\df(\M)$ has the so-called empty team property:
\begin{proposition}\label{empty} For all models $\mA$ and formulas $\phi$ of $\df(\M)$, it holds that $\mA\models_{\emptyset}\phi$. 
\end{proposition}
\begin{proof}
The claim is proved using induction on $\phi$.  
\end{proof}
We also observe that  $\df(\M)$ satisfies the downward closure property (compare to Proposition \ref{Downward closure}).
\begin{proposition}\label{D(M)-Downward closure}Let $\phi$ be a formula of $\df(\M)$, $\mA$  a model, and $Y\subseteq X$ teams. Then $\mA\models_X \phi$ implies $\mA\models_Y\phi$.
\end{proposition}
\begin{proof} The claim is proved using induction on $\phi$. We consider the case where $\phi$ is $\M x\psi$. The other cases are proved exactly as for dependence logic (see Proposition 3.10 in \cite{vaananen2007}).
By the induction assumption, $\psi$ satisfies the claim. Let $\mA$, $X$ and $Y$ be as above and suppose that $|A|=n$, $|X|=m$, and $|Y|=m-1$.  Let us assume $\mA\models_X\phi$. Then for at least $(n^m)/2$ many functions $F\colon X\rightarrow A$ it holds that $\mA\models_{X(F/x)}\psi$. Since $\psi$ satisfies the claim, it holds that if $\mA\models_{X(F/x)}\psi$, then  $\mA\models_{Y(F'/x)}\psi$, where
\begin{equation}\label{reduct}
F'=F\upharpoonright Y.
\end{equation}
Note  that, in the worst case, at most $n$ different functions $F$ gives rise to the same reduct $F'$ in \eqref{reduct}. Therefore, the number of functions $F\colon Y\rightarrow A$ satisfying  $\mA\models_{Y(F/x)}\psi$ is at least $(n^m)/2n=n^{m-1}/2$ and hence $\mA\models_Y\phi$. It is easy to see that the analogous argument can be used with any $Y\subseteq X$.
\end{proof}

A well-studied property in the context of dependence logic is that of \emph{coherence}, defined as follows.
A formula $\phi$ is called $k$-\emph{coherent} if and only if for all structures $\mA$ and teams $X$ it holds that
\[
\mA \models_{X} \phi \Leftrightarrow \text{ for every $k$-element subteam } X' \subseteq X \text{ it holds that } \mA \models_{X'} \phi.
\]
1-coherent formulas are also called \emph{flat}.

\begin{proposition}\label{Incoherence} There is a formula $\phi \in \df(\M)$  without dependence atoms such that $\phi$ is not $k$-coherent for any $k \in \mathbb{N}$.
\end{proposition}

We also note that the analogue of Proposition \ref{freevar} does not hold for $\df(\M)$.

\begin{proposition}\label{Nonrel} The truth of a $\df(\M)$-formula $\phi$ may depend on the interpretations of variables that do not occur free in $\phi$. 
\end{proposition}

Due to space restrictions, the proofs of Propositions \ref{Incoherence} and \ref{Nonrel}  are deferred to the appendix.

\subsection{Second-order Majority Quantifiers and the Counting Hierarchy}

In this section we define the counting hierarchy and the relevant  generalized quantifiers. 

\begin{definition} Let $k\ge 1$. We define the $k$-ary second-order generalized quantifier $\most^k$ binding a $k$-ary relation symbol $X$ in a formula $\phi$. Assume  $\mA$  is a structure with domain $A$ such that $|A|=n$. Then the semantics of this quantifier is defined as follows:
\[ \mA \models \most^k  X\phi(X) \iff \bigl|\bigl\{ B\subseteq A^k\ |\ \mA \models \phi(B) \bigr\}\bigr|\ge 2^{n^k}/2.\]
We will also make use of the so-called $k$-ary second-order \emph{Rescher quantifier}, defined as follows: 
\[ \mA \models \rc^{k} X,Y(\phi(X),\psi(Y)) \iff \bigl|\bigl\{ B\subseteq A^k\ |\ \mA \models \phi(B) \bigr\}\bigr| \ge \bigl|\bigl\{ B\subseteq A^k\ |\ \mA \models \psi(B) \bigr\}\bigr|.\]
\end{definition}

It is quite easy to see that the $\most^k$-quantifier can be defined in terms of the quantifier $\rc^{k}$. In \cite{KontinenJ:Logcch} it was shown that the $k$-ary Rescher quantifier $\rc^k$ can be defined in first order logic with $\most^{k+1}$, and, for $k\ge 2$, already with $\most^{k}$. It is worth noting that in \cite{KontinenJ:Logcch} the quantifiers  $\most^k$ and $\rc^{k}$ are interpreted as strict majority and strict inequality, respectively. All the results of \cite{KontinenJ:Logcch} that we use also hold under the  "non-strict" interpretation adopted in this article.

The  counting hierarchy ($\ch$) is the
analogue  of  the  polynomial  hierarchy,  defined as the oracle hierarchy using as building block probabilistic polynomial time (the class $\pp$) instead of $\np$:
\begin{enumerate}
\item $\CP =\p$,
\item\label{ch} $\Ck1P=\pp^{\CkP}$,
\item $\ch=\bigcup_{k\in \mathbb{N}}\CkP$.
\end{enumerate}
The counting hierarchy was first defined by Wagner \cite{Wagner:comcps} but  
the above equivalent formulation is due to Tor{\'a}n \cite{Toran:Comcdc}.
The  class  $\pp$ consists  of  languages $L$  for  which  there is  a
polynomial time-bounded nondeterministic Turing machine $N$ such that,
for all inputs $x$, $x\in L$ iff more than half of the computations of
$N$ on input $x$ accept.

In \cite{KontinenJ:Logcch} it was shown that the extension $\FO(\most)$ of $\FO$ by the quantifiers $\most^k$, for $k\in \mathbb{N}$,
describes exactly the problems in the counting  hierarchy. The proof therein used the fact that 
the  second-order existential quantifier can be simulated by $\most^k$ and first-order logic. 

\begin{theorem}$\FO(\most)=\SO(\most)=\ch$.
\end{theorem}

By the above remark we see that in the previous theorem the $\most$ quantifiers can be replaced by Rescher quantifiers.

\section{Majority over Functions}

For our main result that compares second-order logic and dependence logic with majority-quantifiers, it turns out to be helpful to consider a version of the $\most$-quantifier that ranges over functions instead of relations.

\begin{definition} Let $k\ge 1$. We define the $k$-ary second-order generalized quantifier $\most^k \func$ binding a $k$-ary function symbol $g$ in a formula $\phi$. Assume  $\mA$  is a structure with domain $A$ such that $|A|=n$. Then
\[ \mA \models \most^k \func g\ \phi(g) \iff \bigl|\bigl\{ f\colon A^k\rightarrow A\ |\ \mA \models \phi(f) \bigr\}\bigr|\ge n^{n^k}/2.\]
\end{definition}
We denote by $\SO(\mostf)$ the extension of $\SO$ by the quantifiers $\most^k\func$ for all $k\ge 1$. The following elementary properties of  
$\SO(\mostf)$ will be useful.

\begin{proposition}\label{Llaws} The following equivalences hold:
\begin{enumerate}
\item $(\phi \vee \most^k\func g\, \psi) \equiv \most^k\func g\,(\phi\vee \psi)$, if $g$ does not appear free in $\phi$,
\item $(\phi \wedge \most^k\func g\, \psi) \equiv \most^k\func g\,(\phi\wedge \psi)$, if $g$ does not appear free in $\phi$.
\end{enumerate}
\end{proposition}

The equivalences of Proposition \ref{Llaws} obviously hold also for the relational majority quantifiers $\most^k$.

The next proposition states the intuitively obvious fact that the extensions of $\SO$
 by the quantifiers  $\most^k$ or alternatively by  $\most^k \func$, for $k\in \mathbb{N}$, are equal in expressive power.    
\begin{proposition}\label{SO(M)=SO(M_f)} $\SO(\most) = \SO(\mostf)$.
\end{proposition}
\begin{proof} We prove the claim  by an  argument analogous to  Theorem 3.4 in \cite{KontinenJ:Logcch}. We will show how to express the quantifier  $\most^k \func$ in the logic  $\SO(\most)$ implying  $\SO(\mostf)\le \SO(\most)$. The converse inclusion is proved analogously. 

 Let us consider a formula of the form $\most^k \func g \phi(g)\in \SO(\mostf)$.  Let $\mA$ be a structure. We may assume that $\mA$ is ordered (we can existentially quantify it) and hence there is a $\FO$-formula $\delta (\overline{x},\overline{y})$ defining the lexicographic ordering of the set $A^{k+1}$. We can construct   a formula
$\chi(X,Y)$ which, for $A_1,A_2\subseteq A^{k+1}$,  defines the lexicographic ordering ($A_1\le_{l} A_2$) of $k+1$-ary relations  induced by $\delta(\overline{x},\overline{y})$.

It is now  fairly straightforward to express 
$\most^k \func g\phi(g)$
in the logic $\SO(\most)$.
Let 
\begin{eqnarray*}
 G&=&\{B\subseteq A^{k+1}\ | \  \textrm{$B$ is the graph of  some $g$ and }  \mA\models \phi(g)\}, \\ 
G^c &=&\{B\subseteq A^{k+1}\ | \  \textrm{$B$ is the graph of some  $g$ and }  \mA\not\models \phi(g)\}.  
\end{eqnarray*}
It now suffices to express $|G|\ge|G^c|$ in the logic  $\SO(\most)$. For a $D\subseteq A^{k+1}$,  define the set  $\text{IS}(D)$ (the ``initial segment'' determined by $D$) by
\[ \text{IS}(D)=\{ D'\subseteq  A^{k+1}|\ D' \notin G\cup G^c \textrm{ and } D'\le_{l} D\}. \]
The condition $|G|\ge|G^c|$ can be now expressed by 
\[ \forall D \big(  |G^c\cup \text{IS}(D)| \ge  2^{n^{k+1}}/2 \Rightarrow |G\cup \text{IS}(D)| \ge  2^{n^{k+1}}/2\big). \]
It is straightforward to express this in the logic  $\SO(\most)$. 
\end{proof}

The following lemma will be needed in the proof of the next proposition.
\begin{lemma}\label{Padd} Let $k\ge 1$. There exists an $\ESO$ sentence $\chi(g)$, where $g$ is $k$-ary, such that for all $\mA$ with domain $|A|=n$, $\chi(g)$ is satisfied by exactly 
 $\lceil n^{n^k}/2\rceil - 2^{n^k-1}$ 
 many $k$-ary functions $g$ none of which is a characteristic function of some $k$-ary relation, i.e., 
$g(\overline{a})\notin\{0,1\}$ for some $\overline{a}\in A^k$ and distinct elements $0$ and $1$.
\end{lemma}

\begin{proof}Without loss of generality, we may assume that  $A=\{ 0,1,\ldots,n-1\}$ and that $\le$ is the canonical ordering of $A$. Let us first consider the case that $|A|=n$ is even. Let $\varphi(U)$, where $U$ is a unary relation symbol, be the sentence
\begin{equation}\label{varphi}
\forall x(U(0)\wedge (U(x)\leftrightarrow \neg U(x+1))).
\end{equation}
Note that there  is a natural bijection between functions $g$ such that $g(\tu 0)\in U$ and functions $h$ satisfying $h(\tu 0)\not\in U$, namely, if  $f\colon A\rightarrow A$ is such that $f(a)=a+1$ if $a$ is even and  $f(a)=a-1$ otherwise, then 
\[F\colon  g\mapsto f\circ g,   \] 
is such a bijection of $k$-ary functions of $A$.

Then, we set $\chi'(g)\equiv \exists U(\varphi(U) \wedge U(g(\tu 0))\wedge \exists \tu x\ g(\tu x)\notin \{ 0,1\})$. The last conjunct eliminates functions that correspond to a  characteristic function of some $k$-ary relation. Over structures with domain of even cardinality, the sentence $\chi'(g)$ satisfies the claim of the lemma. 

Suppose now that $|A|=n>2$ is odd. Let $c=n-1$, i.e., a definable constant from the linear order. 
Define $\varphi'(U)$ as follows:
\begin{equation*}
\forall x(U(0)\wedge \neg U(c) \wedge (x<c-1 \rightarrow (U(x)\leftrightarrow \neg U(x+1))).
\end{equation*}
Let $\psi(g,U)$ be the following formula:
\[
\psi(g,U)\equiv\begin{array}{l}
(\exists \tu x \ U(g(\tu x)) \wedge \forall \tu y < \tu x \ g(\tu y)=c) \vee (\forall \tu x \ g(\tu x)=c).
\end{array}
\]
We then set $\chi''(g)=\exists U(\varphi'(U) \wedge \psi(g,U)\wedge \exists \tu x \ g(\tu x)\notin\{ 0,1\})$. The proof that $\chi''(g)$ realizes a suitable partition of $k$-ary functions can be explain algorithmically
as follows. Formula $\varphi'(U)$ splits the domain into three parts, one containing $c$ only, one containing elements of $U$ and the rest (of size equal to that of $U$). Functions $g$ are then sorted according to whether the first element whose image under $g$ is not $c$ has its image in $U$ or not. At each step $t$, $t\geq 0$, an equal number of function are accepted and rejected and we postpone the decision about functions $g$ such that $g(t)=c$   to the next steps. At the end, only the constant function $g(\tu x)=c$, for all $\tu x<\tu n$ remain. It is put explicitly into the "good" side by the second disjunct of formula $\psi(g,U)$. Note that we are taking half of the $k$-ary functions which are not characteristic functions of $k$-ary relations hence half of the number:
\[n^{n^k} - 2^{n^k} \mbox{ that is } \lceil n^{n^k}/2\rceil - 2^{n^k-1}.\]
The expected formula is now $\chi(g)\equiv (\chi'(g)\wedge \theta_{even})  \vee (\chi''(g)\wedge \theta_{odd})$, where $\theta_{even}$ (respectively  $\theta_{odd}$) is a $\ESO$-sentence expressing that $|A|$ is even (respectively odd).
\end{proof}

The next proposition gives a useful normal form for sentences of the logic $\SO(\mostf)$.

\begin{proposition}\label{SO(Most)normal_form} Every sentence of  $\SO(\mostf)$ is equivalent to a sentence of the form
\[ \exists \overline{h}^1 \most ^{k}\func g_1\cdots\most ^{k}\func g_l\, \exists \overline {h}^2\theta, \]
where the function symbols in $\overline{h}^1$, and $g_i$ for $1\le i\le l$, are $k$-ary ($k\ge 3$), and $\theta$ is a universal first-order sentence. 
\end{proposition}
\begin{proof}
Note that by Proposition \ref{SO(M)=SO(M_f)} it suffices to show that every sentence 
of the logic $\SO(\most)$ can be transformed to this form. The result in \cite{KontinenJ:Logcch} shows (as pointed out in Lemma 10.5 in \cite{KontinenNiemisto}) that, in the presence of built-in relations   $\{<,+,\times \}$, sentences of $\SO(\most)$ can be assumed to have the form   
\begin{equation}\label{f1}
\most ^{i_1}Y_1\cdots\most ^{i_l}Y_l\, \psi, 
\end{equation}
where $\psi$ is first-order. Furthermore,  when  $l$ in \eqref{f1}  is fixed, we get a fragment of $\SO(\most)$ characterizing the $l$th level of $\ch$, i.\,e., the class $\ClP$.

We will next show how to transform any sentence of the form \eqref{f1} to the required form. The first step is to quantify out the built-in relations $\{<,+,\times \}$ to get a sentence of the form
\begin{equation}\label{f4}
\exists X_{<}\exists X_{+}\exists X_{\times} \most ^{i_1}Y_1\cdots\most ^{i_l}Y_l\, \psi^*. 
\end{equation} 
The relations $X_{<}$, $X_{+}$, and $X_{\times}$
 can be axiomatized as part of $\psi^*$ (compare to case 2 of Proposition \ref{Llaws}).
Then we modify the sentence \eqref{f4} to change the arities of all the quantified relations to some big enough $k$. We need only to  replace all occurrences, say  $Y_i(t_1,\ldots,t_{i_j})$,  of the quantified relation symbols in $\psi^*$  by $Y_i(t_1,\ldots,t_{i_j},0,\ldots,0)$. (Note that the needed constant $0$ can be defined using the linear order.) Increasing the arity of the second-order existential quantifiers in \eqref{f4} is clearly unproblematic. For the majority quantifiers $\most^{i_j}$, we note that for any structure $\mA$ of cardinality $n$ and $B\subseteq A^v$, the number of $k$-ary relations $D\subseteq A^k$ such that  
\begin{equation}\label{padding}
\{ \overline{a}\in A^v\ |\  (\overline{a},0,\ldots,0)\in D \}=B  
\end{equation}
is $2^{n^k-n^v}$, which is independent of $B$. Furthermore, obviously the truth of  
$\psi^*$ with respect to a tuple of $k$-ary relations $D_1,\ldots,D_{l+3}$ only depends on whether $\psi^*(B_1,\ldots,B_{l+3})$ holds, where $B_i$ is the restriction
of $D_i$ defined analogously to \eqref{padding}. This fact allows us to increase also the arity of the majority quantifiers without changing the meaning of the sentence \eqref{f4}. 

Let us then show how to transform the relational  quantifiers in \eqref{f4} into function quantifiers. We claim that  it is possible to replace $\psi^*(X_{<},X_{+}, X_{\times},Y_1,\ldots,Y_l)$ by a formula of the form 
\begin{equation}\label{FOpart}
 \theta(\overline{g})\vee ( \forall \overline{x}(\bigwedge_{1\le i\le l} g_i(\overline{x})\in \{0,1\})\wedge  \psi'(g_{<}/ X_{<},g_{+}/X_{+}, g_{\times}/ X_{\times},g_1/Y_1,\ldots,g_l/Y_l)),
\end{equation}
where $\overline{g}=(g_{<},g_{+}, g_{\times},g_1,\ldots,g_l)$, 
the new function symbols are all $k$-ary and $\psi'$ is obtained from $\psi^*$ by substituting subformulas $Z(t_1,\ldots,t_k)$ by the corresponding $g_{(.)}(1_1,\ldots,t_k) = 1$, where $Z\in\{Y_{1},\dots,Y_{l},X_{<},X_{+},X_{\times}\}$. 

The formula $\theta(\overline{g})$ is a $\ESO$-formula that accepts certain dummy functions in order to shift the border of acceptance from $(2^{|\mA|^k})/2$ (half of $k$-ary relations) to 
$|\mA|^{|\mA|^k}/2$ (half of $k$-ary functions). The logical form of  $\theta$ is 
\[\chi(g_1)\vee \chi(g_2)\vee \cdots  \vee \chi(g_l), \]
where $\chi(g)$ is defined in Lemma \ref{Padd}.
Note that we  repeatedly use case 1 of Lemma \ref{Llaws} to gather  all the formulas $\chi(g_i)$ into $\theta$ which is placed after the block of all majority quantifiers. 

To prove the claim we finally transform the formula \eqref{FOpart} into Skolem normal form to get a sentence of the form
\begin{equation}\label{f2}
\exists g_{<}\exists g_{+}\exists g_{\times} \most ^{k}\func g_1\cdots\most ^{k}\func g_l\, \exists \overline {g}\psi', 
\end{equation} 
where $\psi'$ is a universal $\FO$-sentence. 
\end{proof}

\section{$\SO(\most)=\df(\M)$}

In this section we show that the logics $\SO(\mostf)$ (and thus, by the previous section, $\SO(\most)$) and $\df(\M)$ are equivalent with respect to sentences. 

We will first show a compositional  translation mapping formulas of $\df(\M)$ into sentences of $\SO(\mostf)$. This translation is analogous to the translation from $\df$ into $\ESO$ of \cite{vaananen2007}. 

\begin{lemma}\label{D(M)toSO(most)}
Let $\tau$ be a vocabulary. For every $\df(\M)[\tau]$-formula $\phi$ there is a $\tau\cup \{S\}$-sentence $\psi$ of $\SO(\mostf)$  such that for all models $\mA$ and teams $X$ with $\dom(X)=\Fr(\phi)$ it holds that
\[\mA\models_X \phi \iff (\mA,\rel(X))\models\psi . \]
\end{lemma}

\begin{proof} For technical  reasons to be motived shortly, we will actually prove a slightly more general result showing that for every $\df(\M)[\tau]$-formula $\phi$ and every finite set of variables $\{y_1,\ldots,y_n\} \supseteq \Fr(\phi)$  there is a $\SO(\mostf)[\tau \cup S]$-sentence $\psi$ such that for all $\mA$ and teams $X$ with $\dom(X)=\{y_1,\ldots,y_n\}$ it holds that
\[\mA\models_X \phi \iff (\mA,\rel(X))\models\psi . \]
We will prove the claim using induction on the structure of $\df(\M)$-formulas. In the following we write $ \phi(y_1,\ldots,y_n)$ to mean that $\Fr(\phi)\subseteq \{y_1,\ldots,y_n\}$. The quantifiers $\rc^k$ can be uniformly defined in the logic $\SO(\most)$, hence by the results of the previous section, also in $\SO(\mostf)$. Therefore, we may freely use the quantifiers $\rc^k$ in the translation.

Atomic formulas and their negations are translated exactly in the same way as in the analogous translation from $\df$ into $\ESO$ in \cite{vaananen2007}. The cases $\gamma \dfn \exists y_n \phi(y_1,\ldots,y_n)$ and $\gamma \dfn \forall y_n \phi(y_1,\ldots,y_n)$ are also translated as in \cite{vaananen2007}. Suppose then that 
$\gamma \dfn (\phi\vee \psi)(y_1,\ldots,y_n)$ and that $\phi^*(S)$ and $\psi^*(S)$ already exist by induction hypothesis.
We translate $\gamma$ as follows:
\begin{equation}\label{disjunction}
\gamma^*(S):= \exists Y\exists Z(\phi^*(Y/S)\wedge\psi^*(Z/S) \wedge \forall y_1\dots \forall y_n (S(\overline{y})\rightarrow R(\overline{y})\vee T(\overline{y}))).
\end{equation}
Note that $\gamma^*(S)$ is defined as in  \cite{vaananen2007}. The only difference is that in the case of dependence logic the sentence \eqref{disjunction} can be written using a single sentence $\phi^*(S)$ (and $\psi^*(S)$) that translates $\phi$ over teams with domain $\Fr(\phi)$ (see  Proposition \ref{freevar}).  In the case of $\df(\M)$ the behavior of $\phi$ and $\psi$ over teams $X$ with $\dom(X)=\{y_1,\ldots,y_n\}$ does not in general reduce to their behavior over $X\upharpoonright \Fr(\phi)$ and $X\upharpoonright \Fr(\psi)$ (see Proposition \ref{Nonrel}).   Therefore, to formulate the sentence \eqref{disjunction}, we need sentences $\phi^*(S)$ and $\psi^*(S)$ that are correct translations of $\phi$ and $\psi$ with respect to  teams with domain $\{y_1,\ldots,y_n\}$.

The case $\gamma \dfn (\phi\wedge \psi)(y_1,\ldots,y_n)$ is also analogous to  \cite{vaananen2007}. It remains to consider the case where our formula $\gamma$ is of the form 
\begin{align}\label{majquantif}
\gamma \dfn \M y_n \phi(y_1,\ldots,y_n)
\end{align}
and $\phi$ is a formula for which we have already a translation into an $\SO(\mostf)[\tau \cup S]$ sentence $\phi^*(S)$.
We claim that $\gamma$ can be translated as follows: 
\begin{align}\label{SORescher}
\gamma^*(S)&\dfn\rc^n \,Y, Z (\theta_1(Y),\theta_2(Z))
\end{align}
where
\begin{align*}
\theta_1(Y) &\dfn  \phi^*(Y/S) \wedge \forall y_1 \dots \forall y_{n-1} \exists^{=1} y_n Y(\overline y) \wedge \forall y_1 \dots \forall y_{n-1} (\exists y_n Y(\overline y)\leftrightarrow S(y_1,\dots,y_{n-1}))\\
\theta_2(Z) &\dfn \neg \phi^*(Z/S) \wedge \forall y_1 \dots \forall y_{n-1} \exists^{=1} y_n Z(\overline y) \wedge \forall y_1 \dots \forall y_{n-1} (\exists y_n Z(\overline y)\leftrightarrow S(y_1,\dots,y_{n-1})).
\end{align*}
The following equivalence is now obvious for all $\mA$ and $X$: 
\[
\mA \models_{X} \gamma \Leftrightarrow (\mA, \rel(X)) \models \gamma^*(S).
\]
\end{proof}

Next we will show that, for sentences, Lemma \ref{D(M)toSO(most)} can be reversed. 

\begin{lemma}\label{SO->D} 
Let $\tau$ be a vocabulary and $\phi\in \SO(\mostf)[\tau]$. Then there is a sentence $\psi\in  \df(\M)[\tau]$
 such that for all models $\mA$:
\[ \mA\models \phi \iff   \mA\models \psi. \]
\end{lemma}

\begin{proof} 
By Proposition~\ref{SO(Most)normal_form} we may assume that $\phi$ is of the form:
\begin{equation}\label{form}
\exists \overline{h}^1\most ^{k}\func g_1\cdots\most ^{k}\func g_n\, \exists \overline {h}^2\forall x_1\cdots \forall x_m\psi,
\end{equation} 
where the function symbols in $\overline{h}^1$ and $g_1,\ldots, g_n$ are $k$-ary, and  $\psi$ is  quantifier free. Before translating  this sentence into $\df(\M)$, we will first apply certain reductions to it. First of all, we make sure that the functions  $g_i$ have only occurrences of the form $g_i(x_1,\ldots,x_k)$ in $\psi$. We can achieve this by existentially quantifying new names $f_i$ for these symbols and passing on to the sentence 
\begin{equation}\label{transf1}
\exists \overline{h}^1\most ^{k}\func g_1\cdots\most ^{k}\func g_n\, \exists \overline {h}^2\exists f_1\cdots \exists f_n\forall x_1\cdots \forall x_m   (\bigwedge_{1\le j\le n}g_j(x_1,\ldots,x_k)=f_j(x_1,\ldots, x_k)\wedge   \psi^*),
\end{equation}
where  $\psi^*$ is obtained from $\psi$ by replacing all occurrences of $g_i$ by $f_i$ for $1\le j \le n$. Analogously,  we may also assume that  the functions $h$ in $\overline{h}^1$ have only occurrences  $h(x_1,\ldots,x_k)$ in $\psi$. Here $m$ can always be made at least $k$.

The next step is to transform the  quantifier-free part $\psi^*$ to satisfy the condition that for each of the function symbols $h$ in $\overline{h}^2$ (also $f_i$) there is a unique tuple $\overline{x}$ of pairwise distinct variables such that all occurrences of it in $\psi^*$ are of the form $h(\overline{x})$ ($f_i(\overline{x})$). In order to achieve this,  we might have to introduce new existentially quantified functions and also universal first-order quantifiers (see Theorem 6.15 in \cite{vaananen2007}), but the 
quantifier structure of the sentence \eqref{form} does not change.

We will now assume that the sentence \eqref{form} has the properties discussed above:
\begin{enumerate}
\item The function symbols $h\in \overline{h}^1$ and $g_i$ have only occurrences of the form $h(x_1,\ldots,x_k)$ and $g_i(x_1,\ldots,x_k)$ in $\psi$, respectively.
\item For each $h$ in $\overline{h}^2$ ($f_i$, for $1\le i \le n$) there is a unique tuple $\overline{x}$ of pairwise distinct variables such that all occurrences of $h$ in $\psi^*$ are of the form $h(\overline{x})$ ($f_i(\overline{x})$). 
\end{enumerate}
We will now show how the sentence \eqref{form} can be translated into $\df(\M)$. For the sake of bookkeeping, we assume that $\overline{h}^1=h_1\ldots h_p$, $\overline{h}^2=h_{p+1}\ldots h_r$, and that 
$h_i$ appears in $\psi$ only as $h_i(\overline{x}^i)$.  We claim now that the following sentence of $\df(\M)$ is a correct translation for \eqref{form}:
\begin{equation}\label{D(M)-translation}
\forall x_1\cdots \forall x_k \exists y_1\cdots \exists y_p \M z_1\cdots \M z_n \forall x_{k+1}\cdots \forall x_{m} \exists y_{p+1}\cdots \exists y_r(\bigwedge _{p+1\le j\le r}\dep(\overline{x}^i,y_i)\wedge\theta),  
\end{equation}
where $\theta$ is obtained from $\psi$ by replacing all occurrences of the term $g_i(x_1,\ldots,x_k)$ by the variable $z_i$ and, similarly, each occurrence of $h_i(\overline{x}^i)$ by $y_i$. 

Let us then show that the sentence $\phi$ (see \eqref{form}) and sentence~\eqref{D(M)-translation} are logically equivalent. Let $\mA$ be a structure and let ${\bf h}_1,\ldots,{\bf h}_r$ and ${\bf g}_1,\ldots,{\bf g}_n$ interpret the corresponding function symbols. We will show that the following holds:
\begin{equation}\label{induction}
(\mA, \overline{{\bf h}}, \overline{{\bf g}})\models _X \psi \Leftrightarrow \mA\models _{X^*}\theta,
\end{equation}
where $X=\{\emptyset\}(A/x_1)\cdots(A/x_m)$ and 
\begin{eqnarray*}
X^* = \{\emptyset\}(A/x_1)\cdots(A/x_k)(H_1/y_1)\cdots(H_p/y_p) (G_1/z_1)\cdots(G_n/z_n)  
(A/x_{k+1})&\cdots& (A/x_m)\\
  (H_{p+1}/y_1)&\cdots&(H_r/y_r),
\end{eqnarray*}
where the supplement functions $H_i$ and $G_i$ are defined using the functions ${\bf h}_i$ and ${\bf g}_i$ as follows:
\begin{eqnarray*}
H_i(s)&=& {\bf h}_i(s(x_1),\ldots, s(x_k)) \textrm{ for } 1\le i\le p,\\
H_i(s)&=& {\bf h}_i(s(\overline{x}^i)) \textrm{ for } p+1\le i\le r,\\
G_i(s)&=& {\bf g}_i(s(x_1),\ldots, s(x_k)) \textrm { for } 1\le i\le n,
\end{eqnarray*}
and where $s(\overline{x}^i)$ is the tuple obtained by  pointwise application of $s$. The claim in \eqref{induction} is now proved using induction on the structure of the quantifier-free formula $\psi$. 
Note that $\psi$ is a first-order formula of dependence logic; hence, by Theorem \ref{FO}, \eqref{induction} holds iff
the equivalence holds for each $s\in X$ (equivalently $s\in X^*$ since the values of the universally quantified variables functionally determine the values of all the other variables) individually. We can now show, using induction on the construction of $\psi$, that for all $s\in X^*$ it holds that
\begin{equation}\label{newadd}
 \mA\models_s \theta \iff (\mA, \overline{{\bf h}}, \overline{{\bf g}}) \models_{s'} \psi,  
\end{equation}
where $s'=s\upharpoonright \{x_1,\ldots,x_m\}$. The key to this result is the fact that, for every $s$, the interpretation of the variables $z_i$ and $y_i$ agree with the interpretation of the terms  $h_i(\overline{x}^i)$ and $g(x_1,\ldots,x_k)$, respectively.

Finally, we note that there is a one-to-one correspondence between  all possible interpretations  ${\bf h}_1,\ldots,{\bf h}_r$ and ${\bf g}_1,\ldots,{\bf g}_n$ for the function symbols and teams $X^*$ satisfying the dependence atomic formulas in \eqref{D(M)-translation}. Therefore, sentence $\phi$ (see \eqref{form}) and sentence~\eqref{D(M)-translation} are logically equivalent. 
\end{proof}

\section{Conclusion and Open Questions}
We have seen that extending dependence logic by a majority quantifier increases the expressive power of dependence logic considerably. One particular consequence of our result is that $\df(\M)$ is closed under classical negation on the level of sentences.  Note further that, for open formulas, this does not hold because of the downward closure property of formulas.

Several open questions remain and we now discuss some of them. Firstly,  
Proposition \ref{Incoherence} shows that the fragment of $\df(\M)$ without dependence atoms does not satisfy the flatness property. It would be interesting to pin down the exact  expressive power of sentences of $\df(\M)$  without dependence atoms. 

The second open question concerns the open formulas of $\df(\M)$. In \cite{kontinen+2007} it was shown that the open formulas of $\df$ correspond to the downwards monotone properties  of $\np$ (see \cite{kontinen+2007} for the exact formulation). We conjecture that the open formulas of $\df(\M)$ correspond in an analogous manner to the downwards monotone properties of $\ch$. 

The majority quantifier is only one particular example of so-called \emph{generalized quantifiers} (or, \emph{Lindstr\"om quantifiers}), introduced in \cite{LindstromP:Firopl} and studied extensively in the context of descriptive complexity theory
 (see  surveys \cite{Vaananen:Genqi} and \cite{EbbinghausH:Finmt}). In \cite{Burtschick:Linqll}, second-order Lindstr\"om quantifiers were introduced and some results concerning their expressive power were obtained. We consider it an interesting study to enrich in a similar way dependence logic by further generalized quantifiers and relate the obtained logics to those studied in \cite{Burtschick:Linqll}.

\newpage

\section{Appendix}
\begin{proof}[Proof of Proposition \ref{Incoherence}]
At first we give a counterexample to the flatness property, i.e., we will give an example such that $\phi \dfn \M x_3 \; x_2 \neq x_3$ does not hold on a team with at least two assignments but it does hold on every unary subset of this team. 

Let $\mA$ be any structure with domain $A \dfn \{0,1,2\}$ and the team $X$ be defined as in \cref{teamXDMnotflat}.

\begin{table}[ht]
\begin{minipage}[b]{0.5\linewidth}
\begin{center}
\[\begin{array}{|c|c|c|}
\hline
& x_1 & x_2\\
\hline
s_1 & 0 & 0\\
\hline
\end{array}
\]
\caption{The team $X$}
\label{teamXDMnotflat}
\end{center}
\end{minipage}\quad
\begin{minipage}[b]{0.5\linewidth}
\begin{center}
\[\begin{array}{|c|c|c|}
\hline
& x_1 & x_2 \\
\hline
s_1 & 0 & 0 \\
\hline
s_2 & 1 & 0 \\
\hline
\end{array}\]
\caption{The team $Y$}
\label{teamXextDMnotflat}
\end{center}
\end{minipage}
\begin{minipage}[b]{0.5\linewidth}
\begin{center}
\[\begin{array}{|c|c|c|}
\hline
& x_1 & x_2 \\
\hline
s_1 & 0 & 0 \\
\hline
s_2 & 1 & 0 \\
\hline
s_3 & 2 & 0 \\
\hline
\end{array}\]
\caption{The team $Z$}
\label{teamZextDMnotflat}
\end{center}
\end{minipage}
\begin{minipage}[b]{0.5\linewidth}
\begin{center}
\[\begin{array}{|c|c|c|}
\hline
& x_1 & x_2 \\
\hline
s_1 & 0 & 0 \\
\hline
s_2 & 1 & 0 \\
\hline
\vdots & \vdots & \vdots \\
\hline
s_{k+1} & n & 0 \\
\hline
\end{array}\]
\caption{The team $X$ showing that $\phi$ is not $k$-coherent}
\label{teamXextDMnokcoherent}
\end{center}
\end{minipage}
\end{table}

The number of functions $F: |X| \to |A|$ is $3^1 = 3$. Let $A' \dfn \{1,2\}$, then every function $F': X \to A'$ satisfies $\mA \models_{X(F'/x_3)} x_2 \neq x_3$. From $|A'|^{|X|} = 2$ we have $\mA \models_{X} \phi$.

Let us now consider team $Y$ depicted in \cref{teamXextDMnotflat}. By an analogous argument, formula $\phi$ holds in all one element subteams $X$ of $Y$. However, among the $3^2=9$ possible ways of supplementing $Y$, there are $|A'|^{|X|} = 4$ many functions $F': X \to A'$ such that  $\mA \models_{Y(F/x_3)} x_2 \neq x_3$. Thus, $\mA \not\models_{Y} \phi$. Since the assignments of $Y$ agree on the variables considered in $\phi$, we may assume, that $X$ represents every subteam of $Y$.

Furthermore, if we consider \cref{teamXextDMnotflat} over a structure $\mA$ with a domain of size 4 (here $A\dfn \{0,1,2,3\}, A' \dfn \{1,2,3\}$) we have, that $|A|^{|Y|} = 4^2 = 16$. Again have $|A'|^{|Y|} = 3^2=6$ many functions $F': X \to A'$ such that $\mA \models_{X(F'/x_3)} x_2 \neq x_3$. Since there are $|A|^{|X|}=9$ many supplementing function we have $\mA \models_{Y} \phi$.

But if we construct the team $Z$ by adding to $Y$ one more assignment as depicted in \cref{teamZextDMnotflat}, we have that $|A|^{|X|} = 4^3 = 64$ but the number of functions $G: A \to X$ with $\mA \models_{Z(G/x_3)} x_2 \neq x_3$ then is $3^3 = 27$ which is less than $32$. Hence, $\mA \not\models_{Z} \phi$.

We can generalize this observation showing, that $\phi$ is not $k$-coherent
by the following construction. Let $X$ be a team of assignments $\{s_1,\dots, s_k, s_{k+1}\}$, where $s_i(x_1) \dfn i$ and $s_i(x_2) \dfn 0$ and $\mA$ be a structure such that $m = k + 2 \dfn |A|$. Let $\phi \dfn \M x_3 \; x_2 \neq x_3$. Then there are $(m-1)^{k+1}$ functions $F: X \to A$ which satisfy $x_2 \neq x_3$. 
And from $(m-1)^{k+1} \leq \frac{m^{k+1}}{2}$ we conclude that that $\mA \not\models_{X} \phi$.

However, for every $k$-element subteam $X'$ of $X$ we have $(m-1)^k$ many functions satisfying $x_2 \neq x_3$ and by $(m-1)^k > \frac{m^k}{2}$ if follows that $\mA \models_{X'} \phi$.

Note that the team has domain $\{x_1,x_2\}$ and we have one fixed formula $\phi = \M x_3 \; x_2 \neq x_3$ that is not $k$-coherent for every $k$. 
Only the structure and the team varies as depicted in \cref{teamXextDMnokcoherent}.

\end{proof}

\begin{proof}[Proof of Proposition \ref{Nonrel}]
We give a counterexample to the analogue of Proposition \ref{freevar} for $\df(\M)$. Recall that,  for a team $X$ and $V\subseteq \dom(X)$, $X\upharpoonright V$ denotes 
\[X\upharpoonright V \dfn\{s\upharpoonright V \mid s\in X\}.\]
Proposition \ref{freevar} shows that
\begin{equation}\label{inv} 
\mA \models _X\phi \textrm{ if and only if } \mA \models _{X\upharpoonright V} \phi,
\end{equation}
for all $\phi \in \df$, all structures $\mA$, teams $X$, and $V\supseteq \Fr(\phi)$. It turns out that the left-to-right implication in \eqref{inv} remains true also for formulas of $\df(\M)$. This can be proved using induction on $\phi\in \df(\M)$ with the help of the fact that all formulas of $\df(\M)$ satisfy the downward closure property of Proposition \ref{D(M)-Downward closure}. 

We will give a counterexample for the right-to-left implication. Let $A=\{0,1,2\}$ and $Z$ be as defined in \cref{teamZextDMnotflat} above. Suppose also that $V=\{x_2\}$, and $\phi\dfn \M x_3 \dep (x_3)$. Now the team $Z\upharpoonright V$ contains only one assignment $s$ and $s(x_2)=0$. Hence trivially $A \models_{Z\upharpoonright V(F/x_3)} \dep (x_3)$ for all $F\colon Z\upharpoonright V\to A$, and therefore
\[ A \models_{Z\upharpoonright V} \M x_3 \dep (x_3). \]  
On the other hand,  $A \models_{Z(F/x_3)} \dep (x_3)$ holds for three (out of nine possible) functions $F\colon Z\to A$ only, hence 
\[ A \not \models_Z \M x_3 \dep (x_3). \]
\end{proof}

\end{document}